\documentclass[twocolumn,showkeys]{revtex4}
\usepackage[utf8]{inputenc}
\usepackage{graphicx}

\begin{document}

\title{On the structure of relativistic hydrodynamics for hot plasmas}

\author{Pavel A. Andreev}
\email{andreevpa@physics.msu.ru}
\affiliation{Department of General Physics, Faculty of physics, Lomonosov Moscow State University, Moscow, Russian Federation, 119991.}
%\affiliation{Peoples Friendship University of Russia (RUDN University), 6 Miklukho-Maklaya Street, Moscow, 117198, Russian Federation}

\date{\today}

\begin{abstract}
Novel structure for relativistic hydrodynamics of classic plasmas is derived following the microscopic dynamics of charged particles.
The derivation is started from the microscopic definition of concentration.
Obviously, the concentration evolution leads to the continuity equation and gives the definition of particle current.
Introducing no arbitrary functions,
we consider the evolution of current (which does not coincide with the momentum density).
It leads to a set of new function which, to the best of our knowledge, have not been consider in the literature earlier.
One of these functions is the average reverse relativistic (gamma) factor.
Its current is also considered as one of basic functions.
Evolution of new functions appears via the concentration and particle current so the set of equations partially closes itself.
Other functions are presented as functions of basic function as a part of truncation presiger.
Two pairs of chosen functions construct two four vectors.
Evolution of these four vectors leads to appearance of two four tensors which are considered instead of the energy-momentum tensor.
The Langmuir waves
%and the electromagnetic waves propagating parallel to the external magnetic field
are considered within the suggested model.
\end{abstract}

%\pacs{}% PACS, the Physics and Astronomy
                             % Classification Scheme.

\maketitle

%52.30.Ex	Two-fluid and multi-fluid plasmas
%52.35.Dm	Sound waves

%52.27.Ep	Electron-positron plasmas

%%%%%%%%%%TEXT

%\mbox{\boldmath $\sigma$}

\section{Introduction}

The universe contains a variety of objects containing plasmas being in extreme conditions,
where relativistic beams going through the relativistically hot plasmas.
Therefore, it is necessary to have several tools for study of such plasmas.
Hence, a novel hydrodynamic model for the plasmas with relativistically hot electrons is developed in this paper.

Macroscopic behavior of any physical system follows from the microscopic dynamics of atoms, molecules, electrons and ions
(in this discussion we do not look deeper into structure of matter).
While this sentence is rather obvious for any researcher, the majority of justifications/derivetions of the macroscopic models are based
on the substitution
of the actual motion of particles by the
alternative methods of description,
by the introduction of the classical probability, for instance.
Our goal is to trace the microscopic relativistic dynamics of the classic particles on the macroscopic scale.
The construction of the collective functions such as the exact number of particles in the unit of volume (concentration)
leads to the macroscopic description of the system.
They appear as the functions defined on the proper scale
so these functions are the continuous functions of the coordinate.
The microscopic equations of motion of each particle allows to derive the equations for the evolution of the macroscopic functions.
In such approach the macroscopic function are the exact functions of the microscopic motion,
since no averaging in the traditional meaning is made.

Major focus of this work is on plasmas with the relativistically hot electrons.
However, the structure of hydrodynamic equations,
possible functions required for the description of relativistic objects are also discussed.
The formal presentation of the relativistic hydrodynamic is made in the Landau books (see for instance \cite{Landau v6}).
However, if any object liquid/gas goes to the relativistic temperatures $T\sim m_{e}c^{2}$ the ionization happens,
where $m_{e}$ is the mass of electron, $c$ is the speed of light.
So we deal with the plasmas anyway.

Some derivations of relativistic hydrodynamics are based
on the consideration of the macroscopically motionless fluid with further transition to the arbitrary frame via the Lorentz transformation or
corresponding construction of the equations in the Lorentz covariant notations \cite{Landau v6}.
However, if the fluid has the small temperature
no global translation can transform it to the regime of large temperatures.
The Lorentz transformation gives a possibility for the relativistic flux in the system.
So these results correspond to the cold plasmas.
Moreover, if we start our derivation from the microscopic scale
we do not make any suggestion relatively the macroscopic state of plasmas.
So, no "rest" frame is considered
while we deal with the arbitrary inertial frame.
There are derivations of hydrodynamics based on the kinetic equations
\cite{Hakim book Rel Stat Phys}, \cite{Mahajan PoP 2002}, \cite{Hazeltine APJ 2002}.
First, we should mention that the developed below model can be found from the kinetic description as well.
Second. in order to have the microscopically justified model
we would need to present the derivation of the kinetics itself \cite{Kuz'menkov 91}.
So, the derivation would be more complex.
Therefore, we avoid the unnecessary intermediate steps and present the derivation of hydrodynamics directly from microscopic description.

%Instead of forcing our will during chose of fundamental function for the relativistic plasma analysis
%we follow the physics and use function naturally appearing in the model.
Macroscopic description of the many-particle systems requires to choose proper functions.
Usually several "familiar" functions are suggested for the description of relativistic plasmas:
such as the fluid density, energy-momentum (four-momentum) density.
These functions are familiar for the researchers who study plasma.
But there is the question:
are there parameters the best for the plasma description?
Moreover, the process of truncation of the set of hydrodynamic equations includes the transition from the four-momentum to the four-velocity.
So, it is questionable do we need equation for the four momentum in the first place?
Otherwise, the set of moments of the distribution function are considered as the necessary functions.
In our derivation we follow natural structure of hydrodynamic equations which appears during calculations.
So, some additional/novel functions appear.
These functions can be constructed of the distribution function
if the additional reverse relativistic gamma factors are involved.

We suggest to start our analysis from the definition of one macroscopic function.
Simplest of these functions is the distribution
(which has no relation to the probability, but it means the"positions in space")
of particles in the physical space called the concentration or the density of the number of particles
(exact number of particles in the vicinity of each point of space).
We trace its evolution in order to find other macroscopic functions.
Hence, we derive necessary hydrodynamic function instead of suggesting them out of our preferences or earlier experience in the field of physics.
The traditional derivation of the relativistic hydrodynamics of the ideal liquid based on the energy-momentum tensor.
However, our analysis shows that the energy-momentum tensor does not appear during microscopic analysis of relativistic plasma evolution.

Some functions presented below can be considered as unusual functions.
However, there are examples of unusual hydrodynamic models for the relativistic objects.
For instance, the relativistic quantum hydrodynamics obtained by Takabayasi \cite{Takabayasi Dirac eq}
(see also \cite{Asenjo PP 11} for modern presentation) is an example of such models.
Its form hugely differs from nonrelativistic one.
It can be seen from structure of equations and functions involved in it.

Before we present the derivation of our model we should mention the works of Yu. L. Klimontovich.
Who suggested the derivation of hydrodynamic and kinetic based on the classic microscopic distribution composed of the delta functions
\cite{Klimontovich Plasma}, \cite{Klimontovich Dokl 62}
(see also \cite{Weinberg Gr 72}).
However, the method of further averaging of obtained microscopic equations has no proper explicit definition.
Presented here method can be considered as the further development of the Klimontovich approach,
where the method of "averaging" is given explicitly.
The word averaging is presented in commas since no averaging in the traditional meaning is made,
while the presiger of scaling based
on the calculation of exact number of particles and other parameters in the vicinity of each point of space is suggested
\cite{Drofa TMP 96}, \cite{Kuzmenkov CM 15}.

The quantum effects can appear in plasma-like objects at the high densities and low temperatures
\cite{Andreev PRE 15 SEAW}, \cite{Andreev 1510 Spin Current}, \cite{Ekman PRE 17},
\cite{Silenko PRA 08}, \cite{Ruiz 15}, \cite{Andreev PRE 16}, \cite{Koide PRC 13}, \cite{Uzdensky RPP 14}.
However, the increase of density can lead to appearance of the relativistic effects as well.
In this paper we address the high temperature regime,
so the quantum effects are irrelevant.

This paper is organized as follows.
In Sec. II the background of the model and its derivation is discussed.
In Sec. III suggested set of relativistic hydrodynamic equations is presented and described.
In Sec. IV the method of introduction of the velocity field is presented.
In Sec. V the method of derivation of the equations of state required for the truncation is given.
In Sec. VI the spectrum of collective excitations is considered.
In Sec. VII a brief summary of obtained results is presented.

%{\color{blue}}

\section{Background of the model}

The derivation of hydrodynamic model starts from the microscopic motion of relativistic particles described by the
Newton's second law for relativistic particles
$\dot{\textbf{p}}_{i}=e_{i}\biggl[\textbf{E}_{i}+\frac{1}{c}[\textbf{v}_{i}\times\textbf{B}_{i}]\biggr]$,
for each particle of the system $i=\overline{1,N}$.
The momentum $\textbf{p}_{i}$ of i-th particle is $\textbf{p}_{i}=m_{i}\textbf{v}_{i}/\sqrt{1-\textbf{v}_{i}^{2}/c^{2}}$,
where $\textbf{v}_{i}$ is the velocity of i-th particle,
and $c$ is the speed of light.
We consider the dynamics of electrons (the ions are motionless),
so all masses $m_{i}$ and charges $e_{i}$ have same value,
but we keep the indexes,
for the further generalizations on the multi-species systems.
Functions $\textbf{E}_{i}=\textbf{E}(\textbf{r}_{i}(t),t)$ and $\textbf{B}_{i}=\textbf{B}(\textbf{r}_{i}(t),t)$ present
the electromagnetic field acting on the $i$-th particle.
These functions are caused by the surrounding particles.
They also include the external field.

To start the macroscopic analysis
we present the microscopic definition of the concentration of particles $n(\textbf{r},t)$ in the vicinity of arbitrary point of space
in the arbitrary inertial frame \cite{Kuz'menkov 91}, \cite{Drofa TMP 96}, \cite{Kuzmenkov CM 15}, \cite{Andreev PIERS 2012}
\begin{equation}\label{RHD2019ClL concentration definition} n(\textbf{r},t)=\frac{1}{\Delta}\int_{\Delta}d\mbox{\boldmath $\xi$}\sum_{i=1}^{N}\delta(\textbf{r}+\mbox{\boldmath $\xi$}-\textbf{r}_{i}(t)). \end{equation}
The definition of concentration (\ref{RHD2019ClL concentration definition}) contains an integral operator
which count the particles in the vicinity of the point of space.
This number of particles is divided by volume of the vicinity.

The concentration (\ref{RHD2019ClL concentration definition}) is the scalar field in the three dimensional physical space.
Vector $\textbf{r}$ presents the ariphmetization of the physical space in the inertial frame.
Function $\textbf{r}_{i}(t)$ are the coordinates of particles in the chosen inertial frame.
The concentration $n$ is the number of particles in the unit of volume.
We consider the $\Delta$ vicinity of each point of space $\textbf{r}$.
The volume of the vicinity $\Delta$ is chosen to be large enough so
$(n(\textbf{r})-n(\textbf{r}'))/n(\textbf{r})\ll1$ for $\mid\textbf{r}-\textbf{r}'\mid\ll\sqrt[3]{\Delta}$.
For the large volume $\Delta$ we have the large number of particles in the vicinity.
The transition of the center of vicinity in the nearest point leads to change of number of particles on the integral number.
The function $n(\textbf{r},t)$ is close to the continuous function
if this number is small in compare with the number of particles in the vicinity $\Delta$.
The $\Delta$ vicinity of point $\textbf{r}$ contains some number of particles $N(\textbf{r})$.
This number changes at the change of position $\textbf{r}$.
Moreover, this number changes with time due to the motion of particles.
It gives the definition of the concentration $n(\textbf{r},t)=N(\textbf{r},t)/\Delta$.
However, this form contains no information about evolution of the number of particles $N(\textbf{r},t)$.
Hence, we need to represent definition $n(\textbf{r},t)=N(\textbf{r},t)/\Delta$ in more useful form.
We need to create an "operator"
which would count the number of particles in the $\Delta$-vicinity of point $\textbf{r}$ depending on the exact microscopic evolution of each particle.
this integral operator is presented within equation (\ref{RHD2019ClL concentration definition}).

In definition (\ref{RHD2019ClL concentration definition}) we have vector $\mbox{\boldmath $\xi$}$
which scan the $\Delta$-vicinity of point $\textbf{r}$.
Therefore, vector $\textbf{r}+\mbox{\boldmath $\xi$}$ corresponds to any point in the vicinity.
So, there is vector $\mbox{\boldmath $\xi$}$ such as
$\textbf{r}+\mbox{\boldmath $\xi$}=\textbf{r}_{i}(t)$
if $i$-th particle is in the $\Delta$-vicinity of point $\textbf{r}$ at the moment $t$.
Hence, the integral of the corresponding $\delta$-function gives 1.
If the chosen particle is outside the vicinity
we have $\textbf{r}+\mbox{\boldmath $\xi$}-\textbf{r}_{i}(t)\neq 0$ for any $\mbox{\boldmath $\xi$}$.
Consequently, the integral of this $\delta$-function gives $0$.
Summation in equation (\ref{RHD2019ClL concentration definition}) goes up to full number of particles in the system.
Hence, this operator checks the presence of any particle in the $\Delta$-vicinity of each point $\textbf{r}$ for all moments in time.
Integral in equation (\ref{RHD2019ClL concentration definition}) is equal to the number of particles in the $\Delta$-vicinity $N(\textbf{r},t)$.

Moreover, equation (\ref{RHD2019ClL concentration definition}) shows a useful form of the microscopic definition of concentration
which allows to calculate its evolution in accordance with the exact microscopic motion of interacting particles.

%.....................................

Obviously, vicinity of the same volume is chosen for all points of space.
The transition to another inertial frame moving relatively the chosen frame with velocity $\textbf{u}_{0}$ would change relative distances between the particles along with the corresponding change of the vicinity volume.
Hence, same number of particles appears in the vicinity,
but the volume changes originating the well-known change of the concentration $n'=\gamma_{u}n$,
where $n'$ is the concentration in the "moving" frame and $\gamma_{u}=1/\sqrt{1-\textbf{u}_{0}^{2}/c^2}$.
However, we do not make any transition to other inertial frames during derivation and application of suggested model.

%For the transition from the microscopic level on the macroscopic scale we need to introduce first hydrodynamic function.
%The concentration of particle is chosen as the fundamental parameter:

\section{Set of hydrodynamic equations}

Obviously, the evolution obtained from microscopic dynamics of particles leads to the well-known continuity equation:
\begin{equation}\label{RHD2019ClL continuity equation} \partial_{t}n+\nabla\cdot\textbf{j}=0. \end{equation}
The evolution of number density $n$ is given by the current of particles $\textbf{j}$.
Its explicit microscopic form appears during derivation.
All microscopic definitions presented below are expressed via the same operator:
\begin{equation}\label{RHD2019ClL formula for average}\langle ...\rangle\equiv\frac{1}{\Delta}\int_{\Delta}d\mbox{\boldmath $\xi$}
\sum_{i=1}^{N} ... \delta(\textbf{r}+\mbox{\boldmath $\xi$}-\textbf{r}_{i}(t)),\end{equation}
which is replaced by symbol $\langle ...\rangle$ in order to make definitions more accurate.

Therefore, the particle current can be written in a short form: $\textbf{j}=\langle \textbf{v}_{i}(t)\rangle$,
where $\textbf{v}_{i}(t)=d\textbf{r}_{i}(t)/dt$.
Presented definition of current $\textbf{j}$ gives total velocity of all particles being in the delta vicinity over the volume of the vicinity $\Delta$:
$\textbf{j}=\sum_{i\in\Delta}\textbf{v}_{i}(t)/\Delta$.
It can be represented via the average velocity $\textbf{v}=\textbf{j}/n$.
The average velocity is the vector sum of velocities of all particles being in the delta vicinity over the number of particles in the vicinity:
$\textbf{v}=\sum_{i\in\Delta}\textbf{v}_{i}(t)/N(\textbf{r},t)$.
Hence, average velocity is found as arithmetic mean.

Derivation of the continuity equation does not require equation of motion for each particle.
However, it is necessary for derivation of equations for other collective hydrodynamic variables.
To our goal we use the relativistic Newton equation in terms of the velocity evolution \cite{Landau v2}:
\begin{equation}\label{RHD2019ClL Newton eq} \dot{\textbf{v}}_{i}=\frac{e_{i}}{m_{i}}\sqrt{1-\frac{\textbf{v}_{i}^{2}}{c^{2}}} \biggl[\textbf{E}_{i}+\frac{1}{c}[\textbf{v}_{i}\times\textbf{B}_{i}]
-\frac{1}{c^{2}}\textbf{v}_{i}(\textbf{v}_{i}\cdot\textbf{E}_{i})\biggr], \end{equation}
where $\textbf{v}_{i}=\textbf{v}_{i}(t)$,
$\textbf{E}_{i}=\textbf{E}(\textbf{r}_{i}(t),t)$ and
$\textbf{B}_{i}=\textbf{B}(\textbf{r}_{i}(t),t)$ are the electric and magnetic fields
in the point $\textbf{r}_{i}(t)$ and at time $t$ acting on $i$th particle.
The right-hand side of equation (\ref{RHD2019ClL Newton eq}) is obviously appears from the Lorentz force.

The evolution of current of particles has expected structure similar to the classic hydrodynamics
\begin{equation}\label{RHD2019ClL current derivative Euler equation}
\partial_{t}j^{a}+\partial_{b}\Pi^{ab}=F^{a},
\end{equation}
where
$\Pi^{ab}=\langle v_{i}^{a}(t)v_{i}^{b}(t)\rangle$,
and
$F^{a}=\langle \dot{v}_{i}^{a}(t)\rangle $.
Parameters $m$ and $e$ are the mass and charge of particle,
$c$ is the speed of light,
$\delta^{ab}$ is the three-dimensional Kronecker symbol,
$\varepsilon^{abc}$ is the three-dimensional Levi-Civita symbol.
In equation (\ref{RHD2019ClL current derivative Euler equation}) and below we assume the summation on the repeating index
$v^{b}_{s}E_{b}=\sum_{b=x,y,z}v^{b}_{s}E_{b}$.
Moreover, the metric tensor has diagonal form corresponding to the Minkovskii space,
it has the following sings $g^{\alpha\beta}=\{-1, +1, +1, +1\}$.
Hence, we can change covariant and contrvariant indexes for the three-vector indexes: $v^{b}_{s}=v_{b,s}$.
The Latin indexes like $a$, $b$, $c$ etc describe the three-vectors,
while the Greek indexes are deposited for the four-vector notations.
The Latin indexes can refer to the species $s=e$ for electrons or $s=i$ for ions.
The Latin indexes can refer to the number of particle $j$ at the microscopic description.
However, the indexes related to coordinates are chosen from the beginning of the alphabet,
while other indexes are chosen in accordance with their physical meaning.
However, the content of this equation and its physical meaning are different.
Strictly speaking the introduced force field $F^{a}$ is not exactly the force field,
since the force causes the change of momentum $\dot{\textbf{p}}_{i}=\mathcal{F}$.
Moreover, three-tensor $\Pi^{ab}$ is the flux of particle current.
Hence, it is not the momentum flux (but they coincide in nonrelativistic limit).

Next, let us present the particle current evolution equation in the mean-field approximation (the self-consistent field approximation)
revealing the explicit form of right-hand side.
Overall, we find the following equation
\begin{equation}\label{RHD2019ClL current derivative Euler equation explicit}
\partial_{t}j^{a}+\partial_{b}\Pi^{ab}
=\Gamma\cdot E^{a} +\frac{1}{c}\varepsilon^{abc}\cdot\Theta_{a}\cdot B_{c} -\frac{1}{c^{2}}\Xi^{ab}\cdot E_{b},
\end{equation}
where
$\Gamma=\langle \frac{1}{\gamma_{i}}\rangle$,
$\Theta^{a}=\langle \frac{1}{\gamma_{i}}v_{i}^{a} \rangle$,
and
$\Xi^{ab}=\langle \frac{1}{\gamma_{i}}v_{i}^{a}v_{i}^{b} \rangle$
is the set of new functions.

Since we consider
the mean-field (the self-consistent field) approximation
the inner electromagnetic field obeys the Maxwell equations
$ \nabla \cdot\textbf{B}=0$,
\begin{equation}\label{RHD2019ClL rot E and div E} \begin{array}{cc}
\nabla\times \textbf{E}=-\frac{1}{c}\partial_{t}\textbf{B}, & \nabla \cdot\textbf{E}=4\pi(en_{i}-en_{e}),
\end{array}
\end{equation}
and
\begin{equation}\label{RHD2019ClL rot B with time}
\nabla\times \textbf{B}=\frac{1}{c}\partial_{t}\textbf{E}+\frac{4\pi q_{e}}{c}n_{e}\textbf{v}_{e},\end{equation}
where the ions are considered as the motionless positively charged background.

%\begin{equation}\label{RHD2019ClL div B} \nabla \cdot\textbf{B}=0,\end{equation}
%\begin{equation}\label{RHD2019ClL rot E} \nabla\times \textbf{E}=-\frac{1}{c}\partial_{t}\textbf{B},\end{equation}
%\begin{equation}\label{RHD2019ClL div E with time} \nabla \cdot\textbf{E}=4\pi(en_{i}-en_{e}),\end{equation}

In nonrelativistic hydrodynamics the density of momentum is proportional to the current of particles
which can be considered as the density of velocity.
In the relativistic hydrodynamics they are two different functions
$\langle \gamma_{i}v_{i}^{a}\rangle$
and
$j^{a}=\langle v_{i}^{a}\rangle$.
Hence, we have two "Euler" equations in the relativistic regime.
We do not consider the momentum density,
which is given in the majority of papers on the relativistic plasmas.
Hence, equation (\ref{RHD2019ClL current derivative Euler equation explicit}) is,
to some extend, a novel equation.
It shows that the evolution of the current of particles $\textbf{j}$ appears
via the flux of vector $\textbf{j}$ and via the interaction between particles.
the interaction is presented in the mean-field approximation
and expressed via the electric and magnetic fields satisfying the Maxwell equations
(\ref{RHD2019ClL rot E and div E}) and (\ref{RHD2019ClL rot B with time}).
However, the electromagnetic field is coupled with functions $\Gamma$, $\Theta^{a}$, and $\Xi^{ab}$.
Therefore, we need to find equations for these functions.
Our analysis shows that
it is necessary to get equations for $\Gamma$, and $\Theta^{a}$,
while an equation of state can be applied to $\Xi^{ab}$.
Function $\Xi^{ab}$ is the flux of vector function $\Theta^{a}$.
It can be seen from their definitions given after equation (\ref{RHD2019ClL current derivative Euler equation explicit}).
Moreover, this interpretation follows from the equation of evolution of $\Theta^{a}$ given below.

The evolution equation for the Gamma function in the mean-field approximation is
\begin{equation}\label{RHD2019ClL equation of Gamma evolution 2 monopole}
\partial_{t}\Gamma+\partial_{a}\Theta^{a}= -\frac{1}{c^{2}}\textbf{E}\cdot\biggl(\textbf{j}-\frac{1}{c^{2}}\textbf{Q}\biggr),
\end{equation}
where
$Q^{a}=\langle v_{i}^{a}\textbf{v}_{i}^{2}\rangle$.

Finally, we present the evolution equation for the hydrodynamic Theta function
in the monopole approximation of the electromagnetic field
$$\partial_{t}\Theta^{a} +\partial_{b}\Xi^{ab}=\frac{e}{m}E^{a}\biggl[n-\frac{\Pi^{bb}}{c^{2}}\biggr]$$
\begin{equation}\label{RHD2019ClL equation of Theta evolution}
+\frac{e}{mc}\varepsilon^{abc}\biggl[j_{b}-\frac{Q_{b}}{c^{2}}\biggr]B_{c} -\frac{2e}{mc^{2}}E_{b}\biggl[\Pi^{ab}-\frac{L^{abcc}}{c^{2}}\biggr],\end{equation}
where
$L^{abcd}=\langle v_{i}^{a}v_{i}^{b}v_{i}^{c}v_{i}^{d}\rangle$,
$\Pi^{bb}\equiv \Pi^{b}_{b}$ is the trace of tensor $\Pi^{ab}$,
$L^{abcc}=L^{abc}_{c}$ is the partial trace of symmetric tensor $L^{abcd}$ on two indexes.

Moreover, the details of truncation to the meanfield from the general form of equations is also ignored.
Here, we introduce the main concept
while technical details will be published elsewhere.

The particle current evolution equation suggests
that we need evolution equation for the Xi function $\Xi^{ab}$.
However, it is necessary to create a limited set of equations.
Hence, at this stage of the model development find an equation of state for the Xi function $\Xi^{ab}$.

\section{Velocity field in equations for the relativistic hydrodynamics}

Introduce the velocity field in the traditional way as the ratio between the particle current and the concentration $\textbf{v}=\textbf{j}/n$.
Next, we need to recognize the contribution of the velocity field in other hydrodynamic functions.
The velocity field is the local average velocity or in other words it is the average velocity of all particle in the delta vicinity $\Delta$ of point $\textbf{r}$.
Therefore, we can split the velocity of each particle on the average velocity (the velocity field) $\textbf{v}$
and the deviation from the velocity field $\textbf{u}_{i}$ caused by the difference of velocities of particles related to the thermal effects: $\textbf{v}_{i}=\textbf{v}+\textbf{u}_{i}$.
Hence, $\textbf{u}_{i}$ is the local thermal velocity of particles.
The average of the thermal velocity is equal to zero $\langle\textbf{u}_{i}\rangle=0$.

Substitute the decomposition of the velocities in the definition of the hydrodynamic functions.
Start with the current flux
\begin{equation}\label{RHD2019ClL} \Pi^{ab}=\langle v_{i}^{a}v_{i}^{b}\rangle=nv^{a}v^{b}+p^{ab},\end{equation}
which  gives the function $p^{ab}$.
It resemblances the pressure.
However, the pressure is the flux of momentum via the surface.
Here, we have the flux of velocity.
So, it is a different function
which has physical meaning similar to the pressure.
The terms linear on the thermal velocity go to zero.
Next, consider the hydrodynamic Theta function
\begin{equation}\label{RHD2019ClL}\Theta^{a}=\biggl\langle \frac{v_{i}^{a}}{\gamma_{i}}\biggr\rangle=\Gamma v^{a}+t^{a},\end{equation}
where
$t^{a}=\langle\frac{u_{i}^{a}}{\gamma_{i}}\rangle$,
with
\begin{equation}\label{RHD2019ClL}
\gamma_{i}=\frac{1}{\sqrt{1-[\textbf{v}^{2}+2\textbf{v}\cdot\textbf{u}_{i}+\textbf{u}_{i}^{2}]/c^{2}}}.\end{equation}
Function $t^{a}$ can be called the thermal part of the hydrodynamic Theta function
while function $t^{a}$ also contains the velocity field in nonadditive form.
Same is true for the hydrodynamic Gamma function
since "averaged" reverse gamma factor contains both the velocity field and the thermal velocity.

Describe the structure of the hydrodynamic Xi function $\Xi^{ab}$:
\begin{equation}\label{RHD2019ClL}\Xi^{ab}=\biggl\langle \frac{v_{i}^{a}v_{i}^{b}}{\gamma_{i}}\biggr\rangle
=\Gamma v^{a}v^{b}+ v^{a}t^{b} +t^{a}v^{b} +t^{ab},\end{equation}
where
$t^{ab}=\langle\frac{u_{i}^{a}u_{i}^{b}}{\gamma_{i}}\rangle$.

We ready to represent continuity equation and the particle current evolution equation,
but the hydrodynamic Gamma function evolution equation contains vector $Q^{a}$
and the hydrodynamic Theta function evolution equation includes vector $Q^{a}$ and a partial trace of tensor $L^{abcd}$.
Let us present the structure of presented functions:
\begin{equation}\label{RHD2019ClL}Q^{a}=\langle v_{i}^{a}v_{i}^{b}v_{i,b}\rangle=nv^{a}v^{b}v_{b}+v^{a}p^{b}_{b}+2v_{b}p^{ab}+q^{a},\end{equation}
where
\begin{equation}\label{RHD2019ClL}q^{a}=\langle u_{i}^{a}u_{i}^{b}u_{i,b}\rangle,\end{equation}
Similarly find  large expression for
$L^{abcd}=\langle v_{i}^{a}v_{i}^{b}v_{i}^{c}v_{i}^{d}\rangle$, but we do not present it here.
The partial trace of tensor $L^{abcd}$ is a part of obtained hydrodynamic equation.
So, we show it here $L^{abcc}$:
$$L^{abcc}=nv^{a}v^{b}\textbf{v}^{2} +v^{a}v^{b}p^{cc} +2v^{a}v_{c}p^{bc}+2v^{b}v_{c}p^{ac}$$
\begin{equation}\label{RHD2019ClL Labcc via v}
+\textbf{v}^{2}p^{ab} +v^{a}q^{b}+v^{b}q^{a}+2v_{c}q^{abc} +M^{abcc},\end{equation}
where
$q^{abc}=\langle u_{i}^{a}u_{i}^{b}u_{i}^{c}\rangle$.

Obviously, the four momentum can be introduced in this approach: $P^{\alpha}=\{\varepsilon/c,\textbf{P}\}$,
where $\varepsilon=\langle \gamma_{i}m_{i}c^{2}\rangle$
and
$\textbf{P}=\langle \gamma_{i}m_{i}\textbf{v}_{i}\rangle$.
However, they do not appear during our derivation.
So, we do not discuss them.
In the traditional relativistic hydrodynamics
\cite{Mahajan PoP 2002},
\cite{Mahajan PRL 03},
\cite{Mahajan PoP 2011},
\cite{Comisso PRL 14},
\cite{Shatashvili ASS 97},
\cite{Dzhavakhrishvili SP JETP 73}
the four momentum is reduced to the concentration and the velocity field
via corresponding equations of state in order to get closed set of hydrodynamic equations,
where the continuity equation and the Maxwell equations contain the velocity field.
For the cold plasmas we have $\varepsilon=mc^{2}/\Gamma$.
However, this no direct relation between the energy density $\varepsilon$ and $\Gamma$.

We derive the hydrodynamic equations directly from the mechanics describing the microscopic motion.
Some of hydrodynamic models (or some justifications/derivations of the hydrodynamic models) are based on the application of the kinetic equation.
However, the kinetic model itself requires the derivation from the mechanics in the first place.
Therefore, we avoid the application of kinetics as the unnecessary intermediate step.

\section{On the truncation of the obtained equations}

%{\color{blue}}
Calculate required expressions using relativistic equilibrium function
\begin{equation}\label{RHD2019ClL distrib func eq}
f_{0}(p)=\textrm{Z} e^{-\epsilon/T},
\end{equation}
where
\begin{equation}\label{RHD2019ClL normalization of f}
\textrm{Z}=\frac{n}{4\pi m^{2}cTK_{2}(\frac{mc^{2}}{T})},
\end{equation}
with $T$ is the equilibrium temperature in the energy units,
$p$ is the momentum,
$K_{2}(\xi)$ is the second order Macdonald function,
and $\epsilon=\sqrt{m^{2}c^{4}+p^{2}c^{2}}$.

The distribution functions are mostly obtained from the application of the probability theory.
Hence, the distribution functions give the probability density to find a particle with the chosen value of momentum.
Being multiplied by the total number of particles
it shows the probability to find the number of particles with the chosen momentum.
So the integral over some interval gives the probability to find the number of particles with the momentum from the chosen interval.
However, the motion of real particles has no relation to probability.
It is strictly determined by the microscopic equations of motion.
Hence, there is the exact number of particles with the momentum from the chosen interval.
Therefore, the distribution functions show this exact number.
However, any distribution function is obtained under some approximations.
Hence, it shows not exact number of particles,
but a number close to it,
but there is certain uncertainty dictated by the chosen approximation made during truncation.
The probability theory allows to guess or reproduce some distribution functions.
Nevertheless, the distribution function can be interpreted via the exact number of particles
in accordance with the atomic structure of matter and the deterministic motion of particles.
Here, we discuss the theory based on the classical mechanics.
However, the Schrodinger equation gives the deterministic evolution of the wave function.
Hence, this conclusion about meaning of the distribution functions is correct to certain extend for the quantum description as well.

To obtain the closed set of hydrodynamic equations
we need to derive equations of state for functions
$p^{ab}$, $t^{ab}$, $q^{a}$, $q^{abc}$, and $M^{abcd}$.
Moreover, the equilibrium value of function $\Gamma$ should be obtained as well.
The concentration gives the normalization of the distribution function (\ref{RHD2019ClL distrib func eq})
$n=\int f_0(p) d^{3}p $.
Other functions can be represented via the equilibrium distribution function:
$\Gamma_{0}=\int [f_0(p)/\gamma] d^{3}p $,
$p^{ab}=\int v^{a}v^{b}f_0(p) d^{3}p $,
$t^{ab}=\int [v^{a}v^{b}f_0(p)/\gamma] d^{3}p $,
and
$M^{abcd}=\int v^{a}v^{b}v^{c}v^{d}f_0(p) d^{3}p $.
Calculation of functions $p^{ab}$, $t^{ab}$, and $M^{abcd}$ leads to the following representations
$p^{ab}=c^{2}\delta^{ab}\tilde{Z}f_{1}(\beta)/3$,
$t^{ab}=c^{2}\delta^{ab}\tilde{Z}f_{2}(\beta)/3$
$M^{abcd}=c^{4}(\delta^{ab}\delta^{cd}+\delta^{ac}\delta^{bd}+\delta^{ad}\delta^{bc})\tilde{Z}f_{3}(\beta)/15$,
where
$\beta=mc^{2}/T$,
$\tilde{Z}=4\pi Z (mc)^{3}=n\beta K_{2}^{-1}(\beta)$,
\begin{equation}\label{RHD2019ClL f 1} f_{1}(\beta)=\int_{1}^{+\infty}\frac{d x}{x}(x^{2}-1)^{3/2}e^{-\beta x}, \end{equation}
\begin{equation}\label{RHD2019ClL f 2} f_{2}(\beta)=\int_{1}^{+\infty}\frac{d x}{x^{2}}(x^{2}-1)^{3/2}e^{-\beta x}, \end{equation}
and
\begin{equation}\label{RHD2019ClL f 3} f_{3}(\beta)=\int_{1}^{+\infty}\frac{d x}{x^{3}}(x^{2}-1)^{5/2}e^{-\beta x}. \end{equation}
Functions $f_{1}(\beta)$, $f_{2}(\beta)$ and $f_{3}(\beta)$ define the coefficients in the dispersion dependencies.
They are calculated numerically below for the chosen values of temperatures.

%\begin{equation}\label{RHD2019ClL}  \end{equation}

%\begin{equation}\label{RHD2019ClL}  \end{equation}

%\begin{equation}\label{RHD2019ClL}  \end{equation}

\section{Waves in the relativistic plasmas}

After the introduction of the velocity field in the derived hydrodynamic equations and application of the isotropic equations of state for the symmetric tensors involved we find the following representation of hydrodynamic equations.
The continuity equation has the traditional form:
\begin{equation}\label{RHD2019ClL cont via v} \partial_{t}n+\nabla\cdot(n\textbf{v})=0.\end{equation}
The velocity field evolution equation
$$n\partial_{t}v^{a}+n(\textbf{v}\cdot\nabla)v^{a}+\frac{\partial^{a}p}{m}
=\frac{e}{m}\Gamma E^{a}+\frac{e}{mc}\varepsilon^{abc}(\Gamma v_{b}+t_{b})B_{c}$$
\begin{equation}\label{RHD2019ClL Euler for v} -\frac{e}{mc^{2}}(\Gamma v^{a} v^{b}+v^{a}t^{b}+v^{b}t^{a})E_{b}
-\frac{e}{mc^{2}}\tilde{t}E^{a} \end{equation}
includes the flux of the thermal velocities $p$, and the interaction contracted of four terms placed on the right-hand side.
The equation of evolution of the Gamma function
\begin{equation}\label{RHD2019ClL eq for Gamma} \partial_{t}\Gamma+\partial_{b}(\Gamma v^{b}+t^{b})
=-\frac{e}{mc^{2}}n\textbf{v}\cdot\textbf{E}\biggl(1-\frac{1}{c^{2}}\biggl(\textbf{v}^{2}+\frac{5p}{n}\biggr)\biggr),\end{equation}
shows the dynamics caused by the electric field only.
The fourth and final equation is the equation for $t^{a}$
which is a part of Theta function:
$$(\partial_{t}+\textbf{v}\cdot\nabla)t^{a}+\partial_{a}\tilde{t}
+(\textbf{t}\cdot\nabla) v^{a}+t^{a} (\nabla\cdot \textbf{v})$$
$$+\Gamma(\partial_{t}+\textbf{v}\cdot\nabla)v^{a}
=\frac{e}{m}nE^{a}\biggl[1-\frac{\textbf{v}^{2}}{c^{2}}-\frac{3p}{nc^{2}}\biggr]$$
$$+\frac{e}{mc}\varepsilon^{abc}nv_{b}B_{c}\biggl[1-\frac{\textbf{v}^{2}}{c^{2}}-\frac{5p}{nc^{2}}\biggr]
-\frac{2e}{mc^{2}}E^{a}p\biggl[1-\frac{\textbf{v}^{2}}{c^{2}}\biggr]$$
\begin{equation}\label{RHD2019ClL eq for t a} -\frac{e}{mc^{2}}nv^{a}v^{b}E_{b}\biggl[1-\frac{\textbf{v}^{2}}{c^{2}}-\frac{9p}{nc^{2}}\biggr]
+\frac{10e}{3mc^{2}}M E^{a}.\end{equation}
This set of equations
which is used to study the small amplitude waves below.

\subsection{Longitudinal one-dimensional waves in the zero magnetic field regime}

%{\color{blue}}
Consider the relativistically hot plasmas
which has no macroscopic currents in the equilibrium state.
Its equilibrium state can be describe by the relativistic Maxwellian distribution.
Hence, required equilibrium equations of state can be gained from this distribution.
Considering system is described by nonzero equilibrium concentration $n_{0}$.
The equilibrium velocity field $\textbf{v}_{0}$ and the electric field $\textbf{E}_{0}$ are equal to zero.
The pressure tensor $p^{ab}$ and tensor $t^{ab}$ are diagonal tensors:
$p^{ab}=p\delta^{ab}$ and $t^{ab}=\tilde{t}\delta^{ab}$.
The "diagonal" form i assumed for tensor $M^{abcd}$ as well:
$M^{abcd}=M_{0}(\delta^{ab}\delta^{cd}+\delta^{ac}\delta^{bd}+\delta^{ad}\delta^{bc})/3$.
Functions $\Gamma_{0}$, $\textbf{t}_{0}$, $p_{0}$, $\tilde{t}_{0}$, $\textbf{q}_{0}$, $M_{0}$ describing the equilibrium state,
and perturbations $\delta p$, $\delta \tilde{t}$ require some equations of state.
The equilibrium magnetic field is equal to zero and perturbations of the magnetic field are absent.

Next, we present the corresponding linearized equations set.
As usual we start with the linearized continuity equation (\ref{RHD2019ClL continuity equation}):
\begin{equation}\label{RHD2019ClL continuity equation lin 1D}
\partial_{t}\delta n+n_{0}\partial_{x} \delta v_{x}=0. \end{equation}

The linearized equation for the evolution of the velocity field
\begin{equation}\label{RHD2019ClL velocity field evolution equation lin 1D}
n_{0}\partial_{t}\delta v_{x}+\partial_{x}\delta p
=\frac{e}{m}\Gamma_{0} \delta E_{x}-\frac{e}{mc^{2}}\tilde{t}_{0}\delta E^{x}
\end{equation}
shows that the dynamic of velocity is defined by the perturbations of the concentration $\delta n$, pressure and the electric field $\delta E_{x}$.
The scalar pressure can be approximately expressed via the concentration using their equilibrium relation
$p=U_{p}^{2} n$.
Hence, we find
$\delta p=U_{p}^{2} \delta n$.
Moreover, we have the Poisson equation for the electric field $\delta E_{x}$
\begin{equation}\label{RHD2019ClL Poisson equation lin}
\partial_{x}\delta E_{x}=-4\pi \mid e\mid \delta n. \end{equation}

Equations (\ref{RHD2019ClL continuity equation lin 1D}), (\ref{RHD2019ClL velocity field evolution equation lin 1D}),
(\ref{RHD2019ClL Poisson equation lin})
form a closed set of equations leading to the eigenfrequency of the system.

We can consider the linearized equations for $\delta\Gamma$ and $\delta t_{x}$ following from
equations (\ref{RHD2019ClL equation of Gamma evolution 2 monopole}) and (\ref{RHD2019ClL equation of Theta evolution}):
\begin{equation}\label{RHD2019ClL evolution of Gamma lin 1D}
\partial_{t}\delta\Gamma +\Gamma_{0}\partial_{x}\delta v_{x}+\partial_{x}\delta t_{x} =0, \end{equation}
and
$$\partial_{t}\delta t_{x} +\partial_{x}\delta \tilde{t}-\frac{\Gamma_{0}}{n_{0}}\partial_{x}\delta p
+\frac{e}{m}\frac{\Gamma_{0}^{2}}{n_{0}}\delta E_{x}$$
\begin{equation}\label{RHD2019ClL evolution of Theta lin 1D}
=\frac{e}{m}n_{0}\delta E_{x} -\frac{5e}{mc^{2}}p_{0}\delta E_{x} +\frac{10e}{3mc^{2}}M_{0}\delta E_{x}, \end{equation}
where $M_{0}^{xxcc}=(5/3)M_{0}$.
However, they do not give any closed set of equations in this regime.
So, no additional wave solution exist.

Equations (\ref{RHD2019ClL continuity equation lin 1D}),
(\ref{RHD2019ClL velocity field evolution equation lin 1D}),
(\ref{RHD2019ClL Poisson equation lin})
lead to the following dispersion equation
\begin{equation}\label{RHD2019ClL Langmuir wave}
\omega^{2}=\biggl(\frac{\Gamma_{0}}{n_{0}}-\frac{\tilde{t}_{0}}{n_{0}c^{2}}\biggr) \omega_{Le}^{2} +U_{p}^{2}k^{2},
\end{equation}
where
$\omega_{Le}^{2}=4\pi e^{2}n_{0}/m$ is the Langmuir frequency.

Solution (\ref{RHD2019ClL Langmuir wave}) corresponds to the Langmuir wave:
the high-frequency longitudinal wave propagating in plasmas with no external fields.
The coefficient in front of the Langmuir frequency differs from $1$ due to the relativistic effects.
Moreover, this coefficient contains the contribution of functions $\Gamma_{0}$ and $\tilde{t}_{0}$
which should be calculated in order to complete the calculation of frequency.

For $\Gamma_{0}$ we find the following approximate expression using the equilibrium distribution function (\ref{RHD2019ClL distrib func eq})
$\Gamma_{0}=n_{0}K_{1}(mc^{2}/T)/K_{2}(mc^{2}/T)$.
Fraction $\frac{K_{1}}{K_{2}}$ tends to 1 at the small temperatures, since the argument of the Macdonald functions goes to infinity.
Function $\tilde{t}_{0}$ is represented via the characteristic velocity $\tilde{t}_{0}=n_{0}U_{t}^{2}$.
Function $M_{0}$ leads to the third characteristic velocity $M_{0}=n_{0}U_{M}^{4}$.

We can use the characteristic velocities to represent the spectrum (\ref{RHD2019ClL Langmuir wave})
$\omega^{2}=(K_{1}/K_{2}-U_{t}^{2}/c^{2})\omega_{Le}^{2} +U_{p}^{2}k^{2}$.

Before we show the dispersion dependencies of the Langmuir waves for different values of the temperature
let us discuss the contribution of different terms.
We have three terms in equation (\ref{RHD2019ClL Langmuir wave}).
The first and second terms are proportional to the Langmuir frequency square $\omega_{Le}^{2}$.
In the nonrelativistic regime we have the unit coefficient in front of $\omega_{Le}^{2}$.
But the relativistic effects decrease the minimal frequency of the high-frequency longitudinal wave.
So there is a possibility to get to the small frequency limit and necessity to include the motion of ions.

Factor $\Gamma_{0}/n_{0}$ equal to $K_{1}/K_{2}$ is a part of coefficient in front of $\omega_{Le}^{2}$.
It is below unit.
So, as it is mentioned above, we have decrease of minimal frequency from $\omega_{Le}$ to $\sqrt{K_{1}/K_{2}}\omega_{Le}<\omega_{Le}$.
However, we have additional decrease caused by $U_{t}^{2}$.
Here, we present values of coefficients $K_{1}/K_{2}$ and $U_{t}^{2}/c^{2}$ in order to compare their contributions
and necessity of account of the second coefficient $U_{t}^{2}/c^{2}$.
We consider three value of temperature $T_{1}=0.1 mc^{2}$, $T_{2}=mc^{2}$, and $T_{3}=10mc^{2}$.
It gives the following values of the dimensionless parameter $\beta=mc^{2}/T$:
$\beta_{1}=10$, $\beta_{2}=1$, and $\beta_{3}=0.1$.
For $\beta_{2}=1$ we obtain $K_{1}/K_{2}=0.38$ and $U_{t}^{2}/c^{2}=\beta f_{2}/(3 K_{2})=0.1$,
where $K_{1}(1)=0.6$, $K_{2}(1)=1.6$, $f_{2}(1)=0.46$.
Hence the contribution of $U_{t}^{2}$ is of order of quoter of $\Gamma_{0}/n_{0}$.
Hence, it is a noticeable effect.
If we increase the temperature to increase the relativistic effects
we choose $\beta_{3}=0.1$ and get $K_{1}/K_{2}=0.05$ and $U_{t}^{2}/c^{2}=\beta f_{2}/(3 K_{2})=0.02$
where $K_{1}(0.1)=10$, $K_{2}(0.1)=200$, $f_{2}(0.1)=100$.
It shows that the factor in front of the Langmuir frequency square decreases
and the relative role of $U_{t}^{2}$ increases.
For the relatively small relativistic effects $\beta_{1}=10$
we obtain $K_{1}/K_{2}=0.91$ and $U_{t}^{2}/c^{2}=0.07$,
where
$K_{1}(10)=2\times 10^{-5}$, $K_{2}(10)=2.2\times10^{-5}$, $f_{2}(10)=4.2\times10^{-7}$.
Hence, the decrease of temperature from the regime of the relativistic temperatures leads
to the decrease of $U_{t}^{2}$ contribution to the neglegible value.
If the temperature is small in compare with $mc^{2}$
we have the nonrelativistic limit,
where
$K_{1}/K_{2}\rightarrow1$ and $U_{t}^{2}/c^{2}\rightarrow0$.

\begin{figure}
\includegraphics[width=8cm,angle=0]{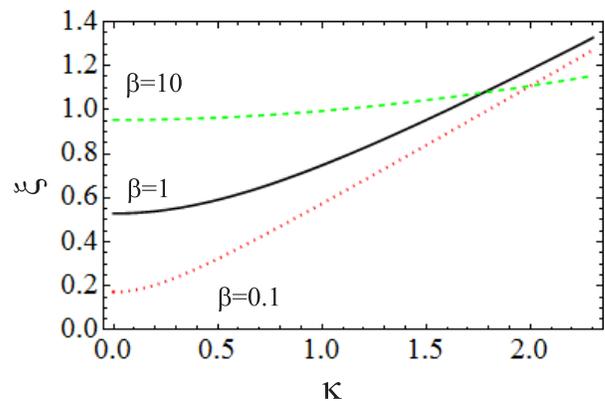}
\caption{\label{RHD2019ClL Fig 01}
The spectrum (\ref{RHD2019ClL Langmuir wave}) is demonstrated numerically in this figure
for three values of temperature $T_{1}=0.1 mc^{2}$, $T_{2}=mc^{2}$, and $T_{3}=10mc^{2}$
which are also discussed in the text.
The figure is presented in terms of the dimensionless variables $\xi=\omega/\omega_{Le}$ and $\kappa=kc/\omega_{Le}$.
We consider the fixed equilibrium concentration $n_{0}$.
Hence, the Langmuir frequency used for the normalization of parameters is fixed either.
}
\end{figure}

The last term in the spectrum shows traditional growth of the frequency at the increase of the wave vector $k$.
However, we should mention that the characteristic velocity $U_{p}^{2}$ related to function $p$
has no direct relation to the pressure (the flux of momentum)
while it presents the flux of velocity (or the flux of the current of particles).
There is no difference between these fluxes in the nonrelativistic limit,
so the obtained spectrum gives the well-known nonrelativistic spectrum.
Nevertheless, $U_{p}^{2}$ is caused by the thermal effects,
so the increase of temperature increases velocity $U_{p}$.
Therefore, the increase of temperature provides
the decrease of the constant part of the spectrum and the increase of the term depending on the wave vector.

%................................................................

The Langmuir wave spectrum can be found from two of four suggested hydrodynamic equations.
However, if we include the magnetic field
we have nonzero contribution of the second term on the right-hand side of equation (\ref{RHD2019ClL Euler for v}).
It contains the perturbation of vector-function $\textbf{t}$.
Hence, the third and fourth equations of the set are involved in the magnetized plasma description.
This comment shows necessity of all four equations.
%%%%%%%%%%%%%%%%%%%%%%%%%%%%%%%%%%%%%%%%%%
However, the regime of magnetized plasmas will be a subject of further papers.

\section{Conclusion}

Novel view on the structure of relativistic hydrodynamics has been suggested.
It evolves through the tracing the microscopic dynamic of particles
which is  averaged over the volume $\Delta$ by the integral operator (see equation (\ref{RHD2019ClL concentration definition})).
Following this dynamics we have arrived to four basic functions $n$, $\textbf{j}$, $\Gamma$, and $\mbox{\boldmath $\Theta$}$
which are parts of two four-vectors $j^{\alpha}=\{n, \textbf{j}\}$, and $\Theta^{\alpha}=\{\Gamma, \mbox{\boldmath $\Theta$}\}$.
Equations are truncated and applied for the Langmuir wave in uniform relativistically hot plasmas.
This regime is chosen for a brief demonstration of thermal relativistic effects following from the suggested model.

\section{Acknowledgements}

Work is supported by the Russian Foundation for Basic Research (grant no. 20-02-00476).
%This paper has been supported by the RUDN University Strategic Academic Leadership Program.

\section{DATA AVAILABILITY}

Data sharing is not applicable to this article as no new data were
created or analyzed in this study, which is a purely theoretical one.

\end{document}